\documentclass[twocolumn, letter]{aa}

\usepackage[T1]{fontenc}
\usepackage{pslatex}
\usepackage{amsmath}
\usepackage{amsfonts}
\usepackage{xcolor}
\usepackage{url}
\usepackage{bm}
\usepackage{graphicx}
\usepackage{amssymb}
\usepackage{soul}
\usepackage{booktabs}
\usepackage{array}
\usepackage[utf8]{inputenc}
\inputencoding{latin1}
\inputencoding{utf8}
\usepackage{appendix}

\def\lsim{ \lower .75ex\hbox{$\sim$} \llap{\raise .27ex \hbox{$<$}} }
\def\gsim{ \lower .75ex \hbox{$\sim$} \llap{\raise .27ex \hbox{$>$}} }

\newcommand{\bi}{\begin{itemize}}
\newcommand{\ei}{\end{itemize}}

\newcommand{\pa}{\partial}

\usepackage[colorlinks=true,linkcolor=blue,citecolor=blue]{hyperref}

\def\simlt{\lower.5ex\hbox{$\; \buildrel < \over \sim \;$}}
\def\simgt{\lower.5ex\hbox{$\; \buildrel > \over \sim \;$}}

\def\pa{\partial}

\newbox\grsign \setbox\grsign=\hbox{$>$} \newdimen\grdimen \grdimen=\ht\grsign
\newbox\simlessbox \newbox\simgreatbox \newbox\simpropbox
\setbox\simgreatbox=\hbox{\raise.5ex\hbox{$>$}\llap
     {\lower.5ex\hbox{$\sim$}}}\ht1=\grdimen\dp1=0pt
\setbox\simlessbox=\hbox{\raise.5ex\hbox{$<$}\llap
     {\lower.5ex\hbox{$\sim$}}}\ht2=\grdimen\dp2=0pt
\setbox\simpropbox=\hbox{\raise.5ex\hbox{$\propto$}\llap
     {\lower.5ex\hbox{$\sim$}}}\ht2=\grdimen\dp2=0pt
\def\simgt{\mathrel{\copy\simgreatbox}}
\def\simlt{\mathrel{\copy\simlessbox}}

\def\cs{c_{\rm s}}

\begin{document}

\title{Noise and quasi-periodic oscillations in black hole accretion flows}
\titlerunning{Noise and QPOs in black hole accretion flows}

\authorrunning{Sobacchi \& Comisso}

\author{
E.~Sobacchi\inst{1,2}
\and
L.~Comisso\inst{3}
}

\institute{
Gran Sasso Science Institute, viale F.~Crispi 7, L’Aquila 67100, Italy\\
\email{emanuele.sobacchi@gssi.it}
\and
INFN -- Laboratori Nazionali del Gran Sasso, via G.~Acitelli 22, Assergi 67100, Italy
\and
Department of Astronomy and Department of Physics, Columbia University, 538 W 120th St, New York, NY 10027, USA
}
\date{}

\voffset-0.4in

\date{}

\abstract{
Temporal fluctuations of the viscosity parameter at different radii in an accretion flow are known to give rise to variability in the accretion rate onto the black hole. The power spectrum of accretion-rate fluctuations is expected to exhibit a broad-band power-law dependence on frequency (e.g.,~flicker noise). We show that viscosity fluctuations can also generate quasi-periodic oscillations (QPOs) in the power spectrum. In our model, QPOs are caused by reflection of accretion-rate fluctuations at the outer radius of the accretion flow and arise near the low-frequency break of the power spectrum, which corresponds to the inverse viscous timescale at the outer radius. These QPOs occur in geometrically thick advection-dominated accretion flows but are absent in geometrically thin cooling-dominated flows.
}


\maketitle
\boldsymbol{}

\section{Introduction}

The X-ray flux from black hole binaries shows significant variability on different timescales. The power spectrum of flux fluctuations has a broad-band power-law dependence on frequency. In addition to broad-band variability, the power spectrum can exhibit quasi-periodic oscillations (QPOs), which appear as narrow-band humps \citep[][]{Vanderklis1989, WijnandsVanderklis1999, IngramMotta2019}. The most common QPOs belong to the type-C class, whose frequency correlates with the low-frequency break of the broad-band power-law. These QPOs are typically detected in the hard X-ray spectral state, which is likely associated with an advection-dominated accretion flow \citep[][]{Done+2007, YuanNarayan2014}.

According to the leading theoretical paradigm for broad-band variability, uncorrelated fluctuations of the viscosity parameter at different radii give rise to a variable accretion rate onto the black hole. When the timescale of viscosity fluctuations is comparable to the local viscous timescale, the power spectrum of the accretion-rate fluctuations follows a broad-band power-law \citep[][]{Lyubarsky1997, Churazov+2001, Kotov+2001, ArevaloUttley2006}. In this scenario, the low-frequency break of the power-law is equal to the inverse viscous timescale at the outer radius of the accretion flow.

One proposed explanation for QPOs is Lense-Thirring precession of the accretion flow \citep[][]{StellaVietri1998, Ingram+2009, IngramDone2011, Musoke+2023}. In this scenario, the broad-band power-law and the QPOs have a different physical origin, making it difficult to explain why the QPO frequency corresponds to the low-frequency break of the power-law. For example, to reproduce the observed correlation between these two frequencies, \citet[][]{IngramDone2011} assumed that the ratio between the Keplerian rotation timescale and the viscous timescale is $0.03r^{-0.5}$, where $r$ is the distance from the black hole in units of Schwarzschild radii. However, such a specific relation is not a general prediction of theoretical models of advection-dominated accretion flows.

In another model \citep[][]{Cabanac+2010}, pressure perturbations in a hot corona are governed by a wave equation. White noise at the outer radius of the corona excites pressure waves. The corona acts as a low-pass medium, generating white noise at low frequencies and red noise at high frequencies. The frequency break  of the power-law corresponds to the inverse wave-crossing time of the corona. Wave reflection at the inner radius produces QPOs near the frequency break. In this model, the origin of white noise at the outer radius is not specified.

In this paper, we investigate whether the broad-band power-law and the QPOs have the same physical origin. We show that fluctuations of the viscosity parameter can generate QPOs. These QPOs are caused by reflection of accretion-rate fluctuations at the outer radius of the accretion flow. Their amplitude and width increase near the low-frequency break of the power spectrum, which corresponds to the inverse viscous timescale at the outer radius. In our model, QPOs arise in advection-dominated accretion flows but are absent when viscous heating is locally balanced by radiative cooling. These properties are qualitatively consistent with the observations of QPOs.

The paper is organized as follows. In Sect.~\ref{sec:model}, we present our model. In Sect.~\ref{sec:results}, we calculate the power spectrum of the accretion rate. In Sect.~\ref{sec:conclusions}, we summarize our conclusions.

\section{Advection-dominated accretion flows with a varying viscosity parameter}
\label{sec:model}

We adopt the slim disk formalism \citep[][]{Abramowicz+1995, Chen+1995}. Assuming vertical hydrostatic equilibrium, the height of the disk is $H=(\cs/v_\phi)R$, where $\cs=\sqrt{P/\rho}$ is the sound speed, and $v_\phi=\sqrt{GM/R}$ is the Keplerian velocity. Here, $P$ and $\rho$ are the pressure and density of the gas, respectively, while $R$ and $M$ are the distance from the black hole and its mass, respectively. The continuity equation gives
\begin{equation}
\label{eq:cont}
\frac{\pa\Sigma}{\pa t} = \frac{1}{2\pi R} \frac{\pa\dot{M}}{\pa R} \;,
\end{equation}
where $\Sigma=\rho H$ is the surface density of the gas, and $\dot{M}=-2\pi R\Sigma v_R$ is the accretion rate, with $v_R < 0$ for inflow. The conservation of angular momentum gives
\begin{equation}
\label{eq:eulerphi}
\frac{\dot{M}}{2\pi} = 3R^{1/2} \frac{\pa}{\pa R} \left( R^{1/2}\nu\Sigma \right) \;.
\end{equation}
The kinematic viscosity coefficient is parametrized as $\nu=\alpha\cs H$, where the viscosity parameter $\alpha$ depends on $R$ and $t$. Since the internal energy per unit area is of the order of $PH$, the energy equation can be written as
\begin{equation}
\label{eq:energy}
\frac{\pa}{\pa t} \left(PH\right) = Q_+ - Q_{\rm adv} \;.
\end{equation}
The viscous heating per unit surface area is given by $Q_+ = 9GM\nu\Sigma/8 R^3$. The advection term is of the order of $Q_{\rm adv}\sim (H/R)^2 Q_+$. We neglected radiative cooling, consistent with our assumption that the flow is advection-dominated.

It is convenient to rewrite Eqs.~\eqref{eq:cont}-\eqref{eq:energy} in terms of the dimensionless variables $\tau=\sigma_{\rm T}\Sigma/m_{\rm p}$, $\dot{m}=\dot{M}/\dot{M}_{\rm edd}$, $h=H/R$, $r=R/R_{\rm s}$, and $s=ct/R_{\rm s}$. Here, $\dot{M}_{\rm edd}=4\pi GMm_{\rm p}/c\sigma_{\rm T}$ is the Eddington accretion rate, and $R_{\rm s}=2GM/c^2$ is the Schwarzschild radius. Eqs.~\eqref{eq:cont} and \eqref{eq:eulerphi} can then be written as
\begin{equation}
\label{eq:taumdot}
\frac{\pa\tau}{\pa s} = \frac{1}{r} \frac{\pa\dot{m}}{\pa r} \;,\qquad
\dot{m} = \frac{3}{\sqrt{2}} r^{1/2} \frac{\pa}{\pa r} \left( \alpha h^2 r \tau \right) \;,
\end{equation}
while Eq.~\eqref{eq:energy} can be written as
\begin{equation}
\label{eq:energydv}
\frac{\pa}{\pa s} \left( h^2\tau \right) = \frac{9\alpha}{8\sqrt{2}} \left(1-\frac{Q_{\rm adv}}{Q_+} \right) \frac{h^2 \tau}{r^{3/2}} \;.
\end{equation}
To recast the evolution equations in a convenient form, we introduce the variables $q=\alpha h^{2}r\tau$ and $x=r^{1/2}$. From Eq.~\eqref{eq:taumdot}, we find
\begin{equation}
\label{eq:final1}
\frac{\partial \tau}{\partial s} = \frac{3}{4\sqrt{2}x^3}  \frac{\partial^2 q}{\partial x^2} \;.
\end{equation}
The accretion rate is given by $\dot{m}=(3/2\sqrt{2})\pa q/\pa x$. Since the ratio $Q_{\rm adv}/Q_+$ is of the order of $h^2=q/\alpha x^2\tau$, we approximate Eq.~\eqref{eq:energydv} as
\begin{equation}
\label{eq:final2}
\frac{\pa q}{\pa s} = \frac{9\alpha}{8\sqrt{2}x^3} \left( 1 - \frac{\xi q}{\alpha x^2\tau} \right) q +\frac{q}{\alpha} \frac{\pa\alpha}{\pa s} \;,
\end{equation}
where $\xi$ is a dimensionless parameter of order unity. For simplicity, we neglect the dependence of $\xi$ on the radial logarithmic derivatives of the flow variables and assume that $\xi$ is constant.

We consider small fluctuations around the stationary solution of Eqs.~\eqref{eq:final1} and \eqref{eq:final2}, writing $q=q_0+q_1$, $\tau=\tau_0+\tau_1$, and $\alpha=\alpha_0+\alpha_1$, where the subscript ``$1$'' indicates the small fluctuations. The stationary solution is $q_0=2\sqrt{2}\dot{m}_0(x-x_{\rm in})/3$ and $\tau_0=\xi q_0/\alpha_0x^2$, where $\alpha_0$ and $\dot{m}_0$ are independent of $x$ and $s$. The unperturbed height of the disk is $h_0=\xi^{-1/2}$. The torque vanishes at the inner boundary of the disk, which is located at $x=x_{\rm in}$. Retaining only linear terms in the small fluctuations, we find
\begin{align}
\nonumber
\frac{\pa^2 q_1}{\pa s^2} & +\frac{9\alpha_0}{8\sqrt{2}x^3} \frac{\pa q_1}{\pa s} = \frac{27\alpha_0^2}{64\xi x^4}\frac{\pa^2 q_1}{\pa x^2} \\
\label{eq:pertaux}
& + \frac{2\sqrt{2}\dot{m}_0\left(x-x_{\rm in}\right)}{3\alpha_0} \frac{\pa^2 \alpha_1}{\pa s^2} +\frac{3\dot{m}_0\left(x-x_{\rm in}\right)}{4x^3}\frac{\pa \alpha_1}{\pa s} \;.
\end{align}
Introducing the variables $X=x/x_{\rm in}$, $S=9\alpha_0s/8\sqrt{2}x^3_{\rm in}$, $Q=3q_1/2\sqrt{2}\dot{m}_0x_{\rm in}$, $a=\alpha_1/\alpha_0$, Eq.~\eqref{eq:pertaux} can be written as
\begin{equation}
\label{eq:pertfinal}
\frac{\pa^2 Q}{\pa S^2} + \frac{1}{X^3} \frac{\pa Q}{\pa S} =
\frac{2}{3\xi X^4}\frac{\pa^2 Q}{\pa X^2} + \left(X-1\right) \left( \frac{\pa^2  a}{\pa S^2} +\frac{1}{X^3}\frac{\pa a}{\pa S}\right) \;.
\end{equation}
Eq.~\eqref{eq:pertfinal} is a telegraph-type equation with a source term proportional to $\pa^2 a/\pa S^2 + X^{-3}\pa a/\pa S$. For $\xi$ of order unity, all terms in Eq.~\eqref{eq:pertfinal} are of the same order when the characteristic timescale of the fluctuations is comparable to the local viscous timescale, which is of the order of $X^3$ in our units.

As we show in Appendix \ref{sec:SSflow} (see, in particular, Eq.~\ref{eq:pertaux2SS}), a telegraph-type equation similar to Eq.~\eqref{eq:pertfinal} can be obtained by considering a cooling-dominated accretion flow and substituting $h_0=\xi^{-1/2}$. In geometrically thin flows where $h_0\ll 1$ (equivalently, $\xi\gg 1$), Eq.~\eqref{eq:pertfinal} correctly describes the evolution of viscous-thermal modes excited by fluctuations of the  viscosity parameter, while in geometrically thick flows, Eq.~\eqref{eq:pertfinal} may be regarded as a phenomenological approximation. To develop a comprehensive theory valid for $\xi\sim 1$, it is essential to account for both the time evolution of angular momentum perturbations and the deviations of the angular velocity from the Keplerian profile \citep[see, e.g.,][]{Kato+1996}. It is also important to use a more accurate model for advection.

\section{Power spectrum of the accretion rate}
\label{sec:results}

We solve Eq.~\eqref{eq:pertfinal} in the radial interval $1<X<X_{\rm out}$. We assume a vanishing torque at the inner boundary of the accretion flow (i.e.,~$Q=0$ at $X=1$) and a constant accretion rate at the outer boundary (i.e.,~$\pa Q/\pa X=0$ at $X=X_{\rm out}$). We adopt the Fourier convention
\begin{align}
\label{eq:fourier1}
Q\left(X,S\right) & = \int_{-\infty}^{+\infty} \frac{{\rm d}\omega}{2\pi} \tilde{Q}\left(X,\omega\right) {\rm e}^{-{\rm i}\omega S} \\
\label{eq:fourier2}
a\left(X,S\right) & = \int_{-\infty}^{+\infty} \frac{{\rm d}\omega}{2\pi} \tilde{a}\left(X,\omega\right) {\rm e}^{-{\rm i}\omega S} \;.
\end{align}
Substituting Eqs.~\eqref{eq:fourier1}-\eqref{eq:fourier2} into Eq.~\eqref{eq:pertfinal}, we find
\begin{equation}
\label{eq:finalfourier}
\left( \omega^2 + \frac{{\rm i}\omega}{X^3} + \frac{2}{3\xi X^4}\frac{\pa^2}{\pa X^2} \right) \tilde{Q} = \left(X-1\right) \left(\omega^2 +\frac{{\rm i}\omega}{X^3}\right) \tilde{a} \;.
\end{equation}
The solution of Eq.~\eqref{eq:finalfourier} can be written as
\begin{align}
\nonumber
\tilde{Q}\left(X,\omega\right) = \int_1^{X_{\rm out}} & {\rm d}Y G\left(X,Y,\omega\right) \\
\label{eq:solution}
& \times \left(Y-1\right) \left(\omega^2 +\frac{{\rm i}\omega}{Y^3}\right) \tilde{a}\left(Y,\omega\right) \;.
\end{align}
To determine the Green's function $G(X,Y,\omega)$ that satisfies the two boundary conditions, we consider the homogeneous equation associated with Eq.~\eqref{eq:finalfourier}, which is
\begin{equation}
\label{eq:hom}
\frac{\pa^2 \tilde{Q}}{\pa X^2} + \frac{3}{2} \xi \omega^2X^4 \tilde{Q} + \frac{3}{2} {\rm i}\xi \omega X \tilde{Q} =0 \;.
\end{equation}
Introducing the variables $Z=X^3/\sqrt{6}$ and $\tilde{U}=Z^{1/3}\tilde{Q}$, Eq.~\eqref{eq:hom} can be written as $\pa^2\tilde{U}/\pa Z^2 = - [\xi\omega^2+{\rm i}\xi\omega/(\sqrt{6}Z) + 2/(9Z^2)] \tilde{U}$. The solution of this equation is a linear combination of the Whittaker functions $M_{\kappa,\;\mu}(-2\sqrt{\xi}{\rm i}\omega Z)$ and $W_{\kappa,\;\mu}(-2\sqrt{\xi}{\rm i}\omega Z)$, where $\kappa=-\sqrt{\xi/24}$ and $\mu=1/6$. The general solution of Eq.~\eqref{eq:hom} is
\begin{align}
\nonumber
\tilde{Q} \left(X,\omega\right) = & \;\frac{C_M}{X} M_{\kappa,\;\mu} \left(-\sqrt{\frac{2\xi}{3}}{\rm i}\omega X^3\right) \\
\label{eq:Qgeneral}
& + \frac{C_W}{X} W_{\kappa,\;\mu} \left(-\sqrt{\frac{2\xi}{3}}{\rm i}\omega X^3\right) \;,
\end{align}
where $C_M$ and $C_W$ are independent of $X$. The solution satisfying the inner boundary condition at $X=1$ (which will be denoted as $\tilde{Q}_{\rm in}$) is obtained for $C_{M,{\rm in}}= W_{\kappa,\;\mu}(-\sqrt{2\xi/3} {\rm i}\omega)$ and $C_{W,{\rm in}}=-M_{\kappa,\;\mu}(-\sqrt{2\xi/3} {\rm i}\omega)$. The solution satisfying the outer boundary condition at $X=X_{\rm out}$ (which will be denoted as $\tilde{Q}_{\rm out}$) is obtained for
\begin{align}
\nonumber
C_{M,{\rm out}} = & \; W_{\kappa,\;\mu} \left(-\sqrt{\frac{2\xi}{3}}{\rm i}\omega X_{\rm out}^3 \right) \\
\label{eq:CMout}
& + \sqrt{6\xi}{\rm i}\omega X_{\rm out}^3 W'_{\kappa,\;\mu} \left(-\sqrt{\frac{2\xi}{3}}{\rm i}\omega X_{\rm out}^3 \right) \\
\nonumber
C_{W,{\rm out}} = & - M_{\kappa,\;\mu} \left(-\sqrt{\frac{2\xi}{3}}{\rm i}\omega X_{\rm out}^3 \right) \\
\label{eq:CWout}
& - \sqrt{6\xi}{\rm i}\omega X_{\rm out}^3 M'_{\kappa,\;\mu} \left(-\sqrt{\frac{2\xi}{3}}{\rm i}\omega X_{\rm out}^3 \right) \;,
\end{align}
where the prime denotes the derivative of the Whittaker functions with respect to their argument. The Green's function can be written as
\begin{align}
\nonumber
G\left(X,Y,\omega\right) = & \; \frac{3\xi Y^4}{2\mathcal{W}\left(\omega\right)} \\
\label{eq:green}
& \times
\begin{cases}
\tilde{Q}_{\rm in} \left(X,\omega\right) \tilde{Q}_{\rm out} \left(Y,\omega\right) & {\rm for}\quad X<Y \\
\tilde{Q}_{\rm in} \left(Y,\omega\right) \tilde{Q}_{\rm out} \left(X,\omega\right) & {\rm for}\quad X>Y
\end{cases}
\;,
\end{align}
where we defined
\begin{equation}
\label{eq:wronskian}
\mathcal{W}\left(\omega\right) = \tilde{Q}_{\rm in} \frac{\pa\tilde{Q}_{\rm out}}{\pa Y} - \frac{\pa\tilde{Q}_{\rm in}}{\pa Y} \tilde{Q}_{\rm out} = - \frac{\pa\tilde{Q}_{\rm in}}{\pa Y} \left(1,\omega\right) \tilde{Q}_{\rm out} \left(1,\omega\right) \;.
\end{equation}
We used the fact that the Wronskian $\mathcal{W}$ is independent of $Y$ because Eq.~\eqref{eq:hom} does not contain first-order derivatives.

\begin{figure}
\centering
\includegraphics[width=0.49\textwidth]{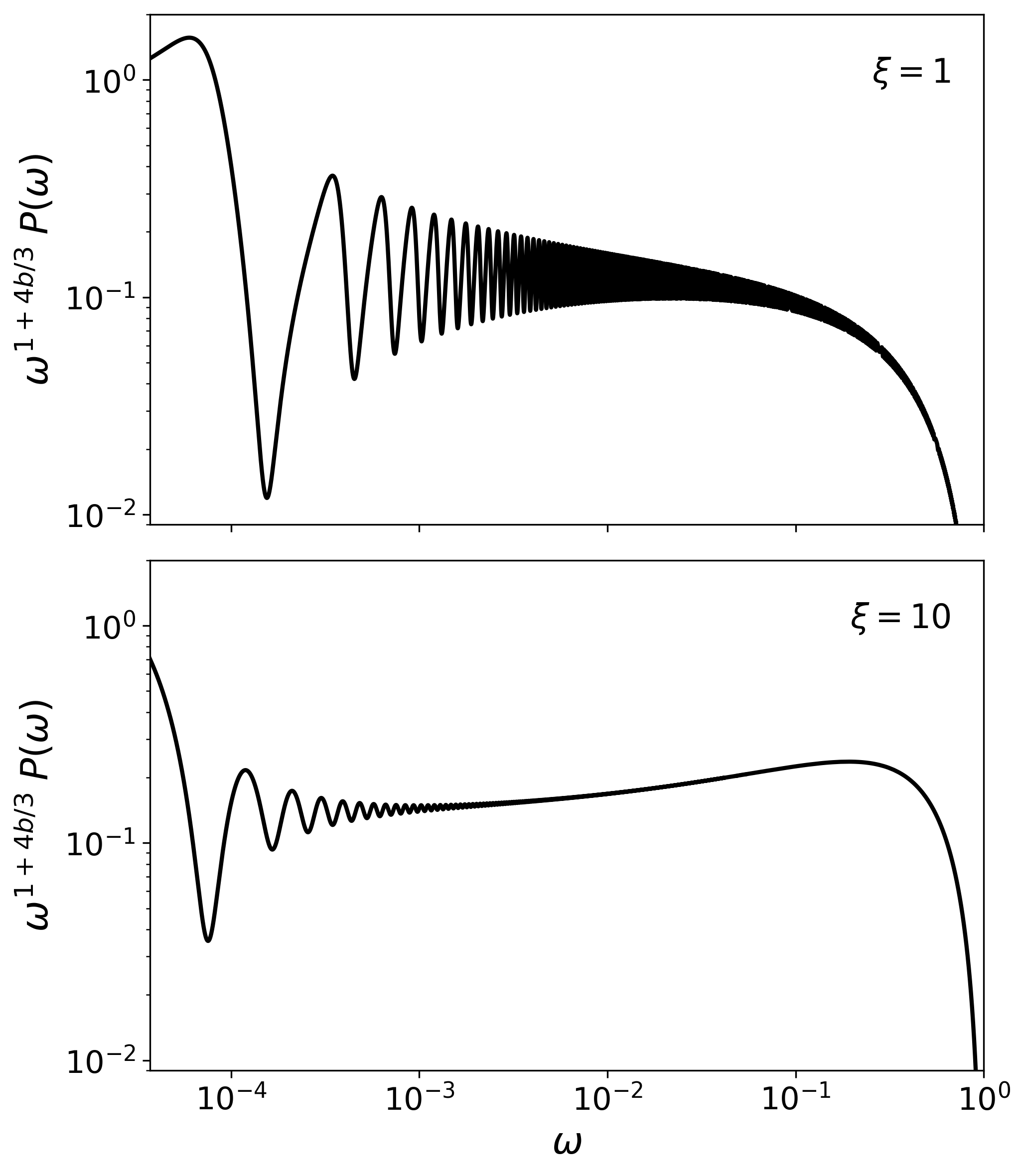}
\caption{\label{fig:power}Power spectrum of the accretion rate onto the black hole for a geometrically thick advection-dominated accretion flow. The power spectrum extends over the frequency range $X_{\rm out}^{-3}<\omega<1$, where $X_{\rm out}=30$ is the dimensionless outer radius of the accretion flow. The top and bottom panels show $\xi=1$ and $\xi=10$ (which correspond to unperturbed dimensionless heights $h_0=1$ and $h_0=0.3$, respectively). The normalization of the power spectrum is arbitrary.}
\end{figure}

In the following, we calculate the power spectrum of the accretion rate at the radius where most of the energy is released, assumed to be located at the inner boundary of the flow (i.e.,~at $X=1$). If most of the energy were released at a larger radius, the low-frequency part of the power spectrum would be unchanged, while the high-frequency break would shift towards lower frequencies \citep[][]{Lyubarsky1997}. The Fourier transform of the fluctuation of the accretion rate in units of $\dot{m}_0$ is given by $\pa \tilde{Q}/\pa X$. Using Eqs.~\eqref{eq:solution} and \eqref{eq:green} to calculate $\pa \tilde{Q}/\pa X$ at $X=1$, we find
\begin{align}
\nonumber
\frac{\pa \tilde{Q}}{\pa X}\left(1,\omega\right) = - \int_1^{X_{\rm out}} & {\rm d}Y \frac{3\xi Y^4}{2} \frac{\tilde{Q}_{\rm out}\left(Y,\omega\right)}{\tilde{Q}_{\rm out}\left(1,\omega\right)} \\
\label{eq:accretion}
& \times \left(Y-1\right) \left(\omega^2 +\frac{{\rm i}\omega}{Y^3}\right) \tilde{a}\left(Y,\omega\right) \;.
\end{align}
The power spectrum of the accretion rate is given by
\begin{align}
\nonumber
& P\left(\omega\right) = \int_1^{X_{\rm out}} {\rm d}Y \int_1^{X_{\rm out}} {\rm d}Y' \frac{9\xi^2Y^4Y'^4}{4} \frac{\tilde{Q}_{\rm out}\left(Y,\omega\right) \tilde{Q}_{\rm out}^*\left(Y',\omega\right)}{\left|\tilde{Q}_{\rm out}\left(1,\omega\right)\right|^2} \\
\label{eq:poweraux}
& \times \left(Y-1\right) \left(Y'-1\right) \left(\omega^2 +\frac{{\rm i}\omega}{Y^3}\right) \left(\omega^2 -\frac{{\rm i}\omega}{Y'^3}\right) C_{\tilde{a}} \left(Y, Y', \omega\right) \;,
\end{align}
where we defined $C_{\tilde{a}}(Y,Y',\omega)=\langle \tilde{a}(Y,\omega) \tilde{a}^*(Y',\omega)\rangle$, and the star denotes the complex conjugate. We assume that (i) the fluctuations of the viscosity parameter at different radii are not correlated; (ii) the timescale of the fluctuations is of the order of the local viscous timescale; (iii) the amplitude of the fluctuations depends on the radius as a power-law of index $b$ \citep[][]{Lyubarsky1997}. For simplicity, we adopt
\begin{equation}
\label{eq:corr}
C_{\tilde{a}} \left(Y, Y', \omega\right) \propto Y^{4b} \delta\left(\log Y-\log Y'\right) \delta\left(\omega-Y^{-3}\right) \;.
\end{equation}
Viscosity fluctuations described by Eq.~\eqref{eq:corr} are completely uncorrelated at different radii, and each radius contributes at exactly one frequency through the delta function. In Appendix \ref{sec:correlation}, we relax these assumptions and show that QPOs can also arise when viscosity fluctuations have a significant correlation length and bandwidth. Substituting Eq.~\eqref{eq:corr} into Eq.~\eqref{eq:poweraux}, we find
\begin{equation}
\label{eq:power}
P\left(\omega\right) \propto \omega^{-1-4b/3} \left(1-\omega^{1/3}\right)^2 \left|\frac{\tilde{Q}_{\rm out}\left(\omega^{-1/3},\omega\right)}{\tilde{Q}_{\rm out}\left(1,\omega\right)}\right|^2 \;.
\end{equation}
Eq.~\eqref{eq:power} gives the power spectrum of the accretion rate in the frequency range $X_{\rm out}^{-3}<\omega<1$. The low- and high-frequency breaks of the power spectrum correspond, respectively, to the inverse viscous timescales at the outer and inner radii of the accretion flow. The factor $\omega^{-1-4b/3}$ shows that the power spectrum depends on the frequency as a power-law \citep[see also Sect.~5 of][]{Lyubarsky1997}. The factor $(1-\omega^{1/3})^2$ smooths the high-frequency break of the power spectrum. The factor $|\tilde{Q}_{\rm out}(\omega^{-1/3},\omega)/ \tilde{Q}_{\rm out}(1,\omega)|^2$ generates the QPOs shown in Fig.~\ref{fig:power}. The amplitude and width of QPOs increase near the low-frequency break of the power spectrum. These properties are similar for $\xi=1$ and $\xi=10$ (which correspond to heights $h_0=1$ and $h_0=0.3$, respectively), and therefore may be independent of details of the accretion flow structure.

\section{Conclusions}
\label{sec:conclusions}

We calculated the power spectrum of the accretion rate onto the black hole in advection-dominated accretion flows with a varying viscosity parameter. The power spectrum has a broad-band power-law dependence on frequency. Quasi-periodic oscillations (QPOs) arise near the low-frequency break of the power spectrum, which corresponds to the inverse viscous timescale at the outer radius of the accretion flow.

The physical origin of QPOs in our model can be understood as follows. A varying viscosity parameter generates accretion-rate fluctuations that propagate through the accretion flow. The frequency and wavelength of these fluctuations depend on the radius at which they are generated. Outward-propagating fluctuations are reflected at the outer radius of the accretion flow and can interfere with inward-propagating fluctuations of the same frequency. The QPOs arise when the interference is constructive. In Appendix \ref{sec:BC}, to support our interpretation of the origin of QPOs, we show that reflection of accretion-rate fluctuations at the outer radius is necessary to generate them, while reflection at the inner radius is not required.

In our model, these QPOs arise only in geometrically thick advection-dominated flows, where accretion-rate fluctuations are governed by a telegraph-type equation. In Appendix \ref{sec:SSflow}, we show that QPOs do not appear in geometrically thin cooling-dominated flows, where the fluctuations are instead governed by a diffusion-type equation. In this regime, outward-propagating accretion-rate fluctuations are strongly damped, preventing the interference required to generate QPOs in our model.

Our model has some limitations that could be addressed in future work. (i) We assumed that the ratio of the advection term to the heating term in the energy equation is proportional to the square of the height of the accretion flow, neglecting the radial logarithmic derivatives of the flow variables. We also prescribed a Keplerian angular velocity and neglected the angular momentum dynamics.
(ii) We assumed a constant accretion rate at the outer radius of the accretion flow, implying that accretion-rate fluctuations are reflected at the outer radius. More realistic boundary conditions, which would require a better theoretical understanding of the interaction between the inner advection-dominated flow and the outer truncated disk, could refine the description of the reflection of accretion-rate fluctuations.
These extensions could further constrain the dependence of QPO properties on the structure of the accretion flow.

\begin{acknowledgements}
We thank the anonymous referee for their detailed and constructive comments that helped improve the paper. We acknowledge insightful discussions with Pasquale Blasi, Yuri Lyubarsky, Gr\'{e}goire Marcel, Gor Oganesyan, Lorenzo Sironi, and Lizhong Zhang. ES is supported by a Rita Levi Montalcini fellowship. LC is supported by NSF grant PHY-2308944 and NASA ATP grant 80NSSC24K1230.
\end{acknowledgements}

\bibliographystyle{aa}
\bibliography{2d}

\appendix

\section{Effect of boundary conditions}
\label{sec:BC}

\begin{figure}
\centering
\includegraphics[width=0.49\textwidth]{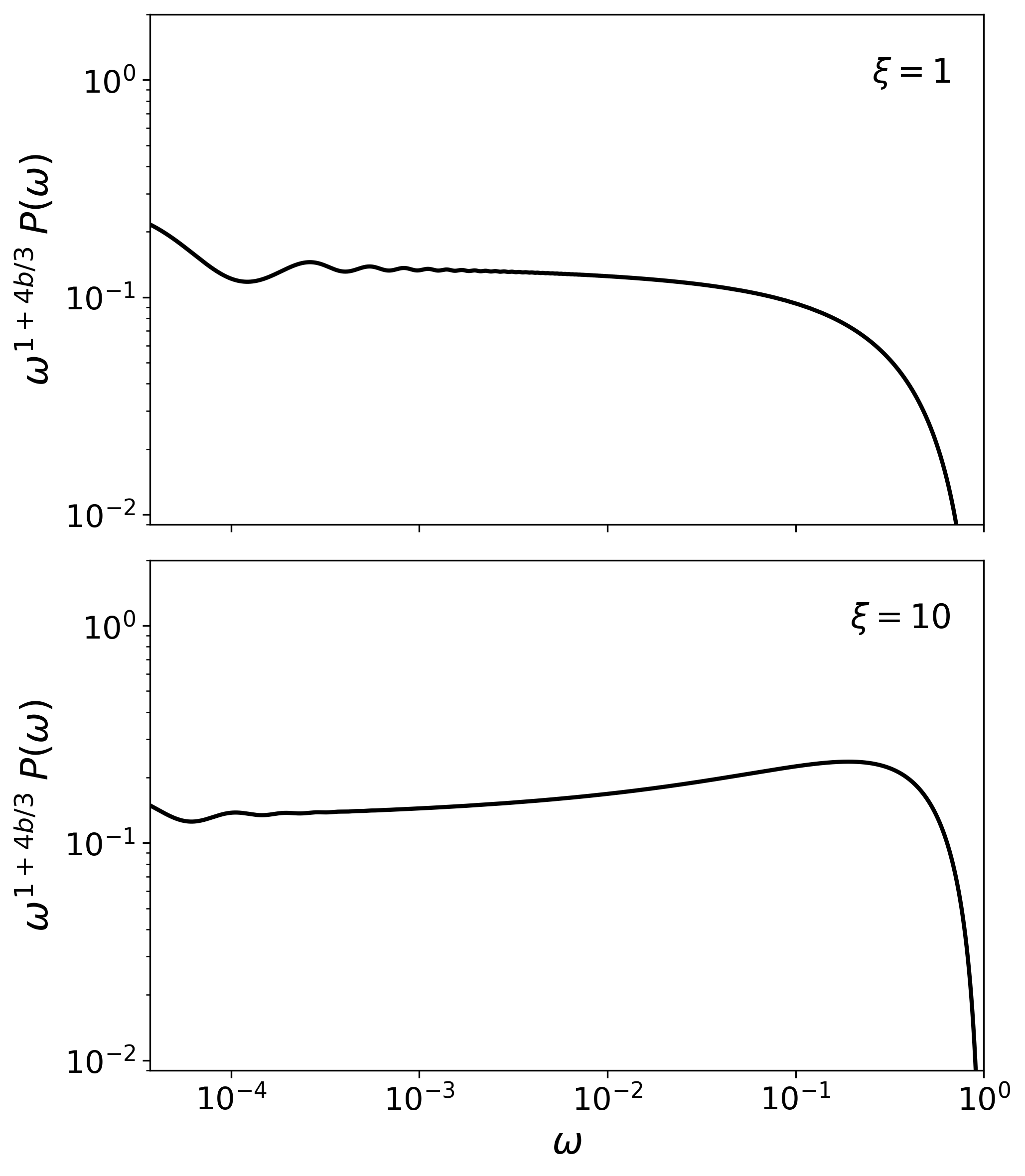}
\caption{\label{fig:BC1}Same as Fig.~\ref{fig:power}, using a local outgoing-wave boundary condition at $X=X_{\rm out}$.}
\end{figure}

We discuss the effect of boundary conditions on the power spectrum of the accretion rate. First, we consider a local outgoing-wave boundary condition at $X=X_{\rm out}$. We assume $\tilde{Q}=0$ at $X=1$, and $\pa\tilde{Q}/\pa X={\rm i}K(X)\tilde{Q}$ at $X=X_{\rm out}$, where we defined
\begin{equation}
K^2 \left(X\right) = \frac{3\xi X^4}{2}\left(\omega^2+\frac{{\rm i}\omega}{X^3}\right)\;.
\end{equation}
The coefficients of the function $\tilde{Q}_{\rm out}$ are given by
\begin{align}
\nonumber
C_{M,{\rm out}} = & \; \left(1+{\rm i}K_{\rm out}X_{\rm out}\right) W_{\kappa,\;\mu} \left(-\sqrt{\frac{2\xi}{3}}{\rm i}\omega X_{\rm out}^3 \right) \\
& + \sqrt{6\xi}{\rm i}\omega X_{\rm out}^3 W'_{\kappa,\;\mu} \left(-\sqrt{\frac{2\xi}{3}}{\rm i}\omega X_{\rm out}^3 \right) \\
\nonumber
C_{W,{\rm out}} = & - \left(1+{\rm i}K_{\rm out}X_{\rm out}\right) M_{\kappa,\;\mu} \left(-\sqrt{\frac{2\xi}{3}}{\rm i}\omega X_{\rm out}^3 \right) \\
& - \sqrt{6\xi}{\rm i}\omega X_{\rm out}^3 M'_{\kappa,\;\mu} \left(-\sqrt{\frac{2\xi}{3}}{\rm i}\omega X_{\rm out}^3 \right) \;,
\end{align}
where we defined $K_{\rm out}=K(X_{\rm out})$. The power spectrum of the accretion rate is given by Eq.~\eqref{eq:power}. As shown in Fig.~\ref{fig:BC1}, QPOs are strongly suppressed, indicating that reflection of accretion-rate fluctuations at the outer radius of the flow is necessary to generate QPOs.

\begin{figure}
\centering
\includegraphics[width=0.49\textwidth]{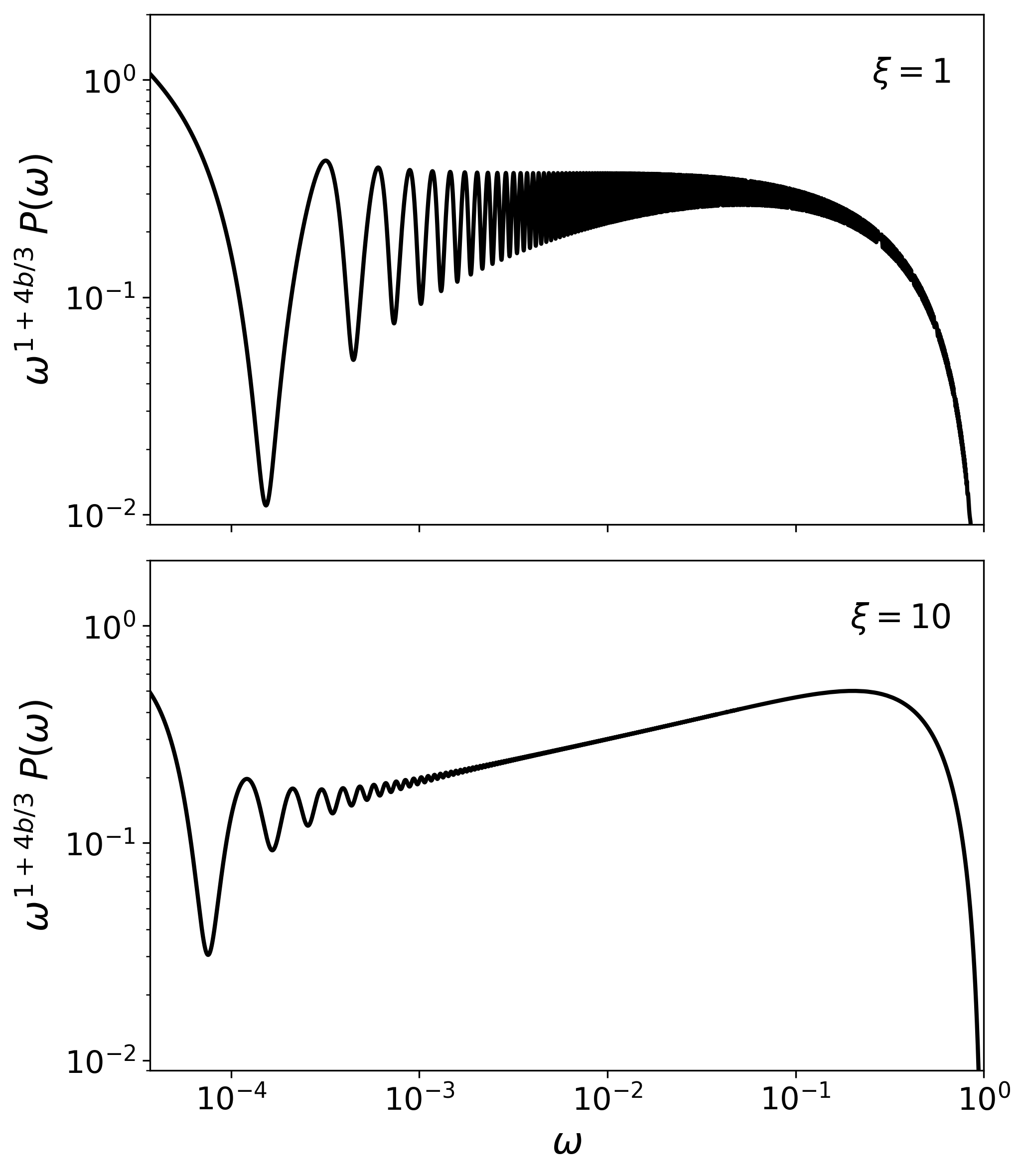}
\caption{\label{fig:BC2}Same as Fig.~\ref{fig:power}, using a local outgoing-wave boundary condition at $X=1$.}
\end{figure}

Now, we consider a local outgoing-wave boundary condition at $X=1$. We assume $\pa\tilde{Q}/\pa X=-{\rm i}K(X)\tilde{Q}$ at $X=1$, and $\pa\tilde{Q}/\pa X=0$ at $X=X_{\rm out}$. The coefficients of the function $\tilde{Q}_{\rm in}$ are given by
\begin{align}
\nonumber
C_{M,{\rm in}} = \left(1-{\rm i}K_{\rm in}\right) & W_{\kappa,\;\mu} \left(-\sqrt{\frac{2\xi}{3}}{\rm i}\omega \right) \\
& + \sqrt{6\xi}{\rm i}\omega W'_{\kappa,\;\mu} \left(-\sqrt{\frac{2\xi}{3}}{\rm i}\omega \right) \\
\nonumber
C_{W,{\rm in}} = - \left(1-{\rm i}K_{\rm in}\right) & M_{\kappa,\;\mu} \left(-\sqrt{\frac{2\xi}{3}}{\rm i}\omega \right) \\
& - \sqrt{6\xi}{\rm i}\omega M'_{\kappa,\;\mu} \left(-\sqrt{\frac{2\xi}{3}}{\rm i}\omega \right) \;,
\end{align}
where we defined $K_{\rm in}=K(1)$. The coefficients of the function $\tilde{Q}_{\rm out}$ are given by Eqs.~\eqref{eq:CMout}-\eqref{eq:CWout}. To calculate the power spectrum, we use the same procedure as in Sect.~\ref{sec:results}. The main difference is that the Wronskian is not given by Eq.~\eqref{eq:wronskian} because $\tilde{Q}_{\rm in}$ does not vanish at $X=1$. Calculating the Wronskian at $X=X_{\rm out}$, we obtain
\begin{equation}
\mathcal{W}\left(\omega\right) = - \frac{\pa\tilde{Q}_{\rm in}}{\pa Y} \left(X_{\rm out},\omega\right) \tilde{Q}_{\rm out} \left(X_{\rm out},\omega\right) \;.
\end{equation}
The power spectrum can be written as
\begin{align}
\nonumber
P\left(\omega\right) \propto & \; \omega^{-1-4b/3} \left(1-\omega^{1/3}\right)^2 \\
\label{eq:powerin}
& \times \left|\frac{\tilde{Q}_{\rm out}\left(\omega^{-1/3},\omega\right)}{\tilde{Q}_{\rm out}\left(X_{\rm out},\omega\right)}\right|^2 \;
\left|\frac{\frac{\pa\tilde{Q}_{\rm in}}{\pa X} \left(1,\omega\right)}{\frac{\pa\tilde{Q}_{\rm in}}{\pa X} \left(X_{\rm out},\omega\right)}\right|^2 \;.
\end{align}
As shown in Fig.~\ref{fig:BC2}, QPOs remain present, indicating that the reflection of accretion-rate fluctuations at the inner radius of the flow is not necessary to generate them. A different choice of inner boundary condition is unlikely to alter this conclusion. Such a choice only modifies the function $\tilde{Q}_{\rm in}$, while an inspection of the terms in Eq.~\eqref{eq:powerin} reveals that QPOs are primarily driven by the function $\tilde{Q}_{\rm out}$.

\section{Effect of correlation of viscosity fluctuations}
\label{sec:correlation}

\begin{figure}
\centering
\includegraphics[width=0.49\textwidth]{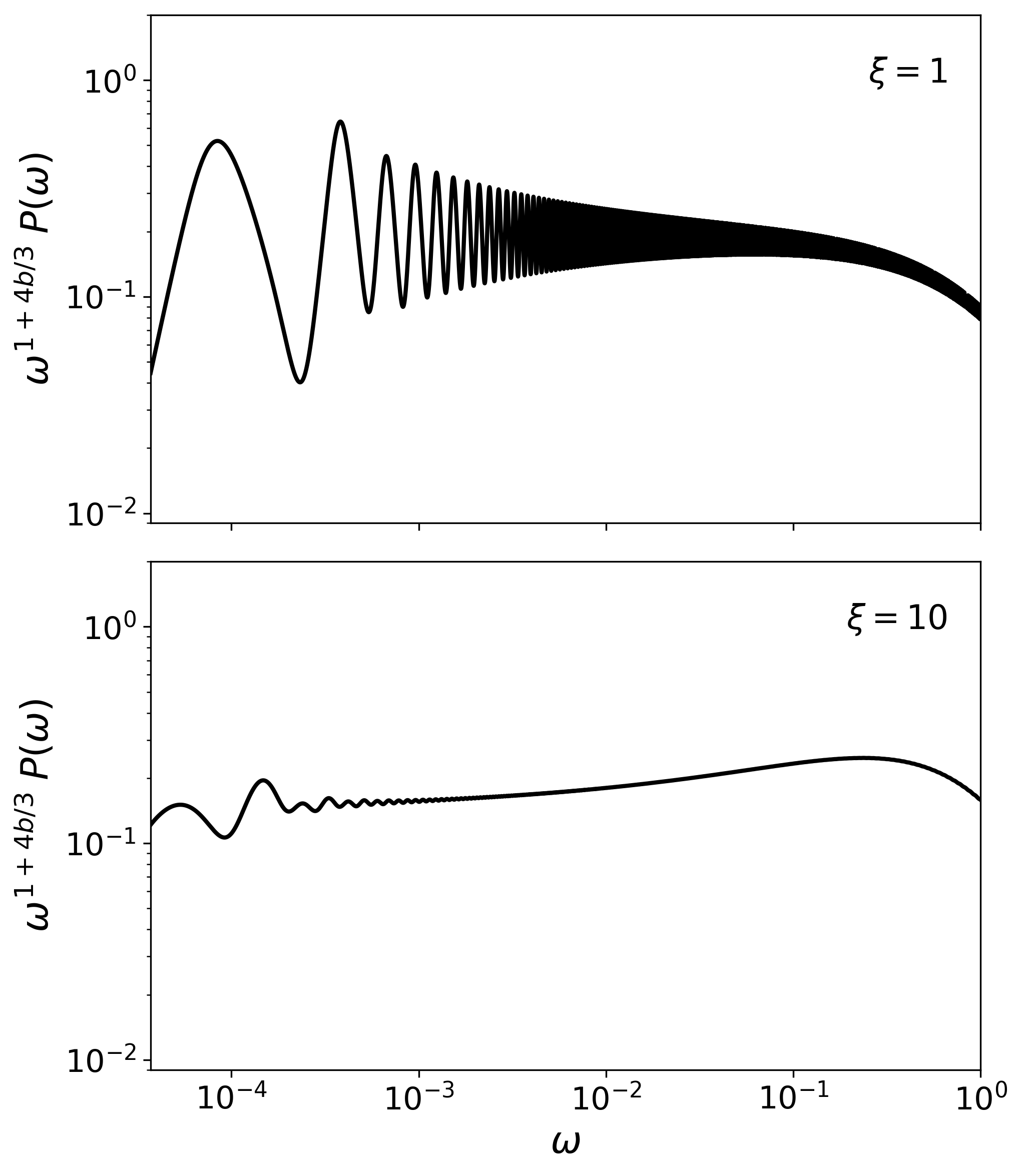}
\caption{\label{fig:powercorr}Same as Fig.~\ref{fig:power}, assuming that the correlation of viscosity fluctuations is described by Eq.~\eqref{eq:corr2}. We adopted $b=0$.}
\end{figure}

In Sect.~\ref{sec:results}, we assumed that viscosity fluctuations are completely uncorrelated at different radii, and each radius contributes at exactly one frequency through the delta function. Here, we relax these assumptions and replace the delta functions with a finite correlation length and bandwidth. We assume that the correlation length and bandwidth are, respectively, of the order of the local radius and inverse viscous timescale. The correlation of viscosity fluctuations is described by
\begin{align}
\nonumber
C_{\tilde{a}} \left(Y, Y', \omega\right) \propto & \left(YY'\right)^{2b} \\
\label{eq:corr2}
\times & \sqrt{S_Y\left(\omega\right) S_{Y'}\left(\omega\right)} \exp \left[-\log^2\left(\frac{Y}{Y'}\right)\right] \;,
\end{align}
where we defined
\begin{equation}
S_Y\left(\omega\right) = \frac{1}{\omega} \exp \left[-\log^2\left(\omega Y^3\right)\right] \;.
\end{equation}
We calculate the power spectrum by substituting Eq.~\eqref{eq:corr2} into Eq.~\eqref{eq:poweraux}. As shown in Fig.~\ref{fig:powercorr}, the properties of QPOs remain qualitatively similar to those presented in Sect.~\ref{sec:results}.

\section{Cooling-dominated accretion flows}
\label{sec:SSflow}

We consider a geometrically thin accretion flow in which viscous heating is locally balanced by radiative cooling, and advection can be neglected \citep[][]{ShakuraSunyaev1973}. The conservation of mass and angular momentum is the same as in advection-dominated accretion flows and is governed by Eqs.~\eqref{eq:cont}-\eqref{eq:eulerphi}. The conservation of energy is, instead, governed by
\begin{equation}
\label{eq:energySS}
\frac{\pa}{\pa t} \left(PH\right) = Q_+ - Q_- \;,
\end{equation}
where $Q_+ = 9GM\nu\Sigma/8 R^3$ and $Q_-=4\sigma T^4/3\tau$. We assume that the gas pressure dominates the radiation pressure, which implies $T=m_{\rm p}P/k\rho$. Then, Eq.~\eqref{eq:energySS} can be written as
\begin{equation}
\label{eq:final2SS}
\frac{\pa q}{\pa s} = \frac{9\alpha q}{8\sqrt{2}x^3} - \frac{\eta mq^4}{\alpha^3\tau^5x^{12}} + \frac{q}{\alpha} \frac{\pa\alpha}{\pa s} \;,
\end{equation}
where $\eta=GM_\odot m_{\rm p}^3c^3\sigma\sigma_{\rm T}/3k^4=6\times 10^{20}$ and $m=M/M_\odot$.

We consider small fluctuations around the stationary solution of Eqs.~\eqref{eq:final1} and \eqref{eq:final2SS} writing $q=q_0+q_1$, $\tau=\tau_0+\tau_1$, $\alpha=\alpha_0+\alpha_1$, where $q_0=2\sqrt{2}\dot{m}_0(x-x_{\rm in})/3$ and $\tau_0=(8\sqrt{2}\eta mq_0^3/9\alpha_0^4x^9)^{1/5}$. Retaining only linear terms in the small fluctuations, we find
\begin{align}
\nonumber
\frac{\pa^2 q_1}{\pa s^2} & + \frac{27\alpha_0}{8\sqrt{2}x^3} \frac{\pa q_1}{\pa s} = \frac{135\alpha_0^2h_0^2}{64x^4}\frac{\pa^2 q_1}{\pa x^2} \\
\label{eq:pertauxSS}
& + \frac{2\sqrt{2}\dot{m}_0\left(x-x_{\rm in}\right)}{3\alpha_0} \frac{\pa^2 \alpha_1}{\pa s^2} +\frac{3\dot{m}_0\left(x-x_{\rm in}\right)}{x^3}\frac{\pa \alpha_1}{\pa s} \;,
\end{align}
where the height of the disk is given by $h_0^2=q_0/\alpha_0x^2\tau_0$. Introducing the variables $X=x/x_{\rm in}$, $S=27\alpha_0s/8\sqrt{2}x^3_{\rm in}$, $Q=3q_1/2\sqrt{2}\dot{m}_0x_{\rm in}$, $a=\alpha_1/\alpha_0$, Eq.~\eqref{eq:pertauxSS} can be written as
\begin{equation}
\label{eq:pertaux2SS}
\frac{\pa^2 Q}{\pa S^2} + \frac{1}{X^3} \frac{\pa Q}{\pa S} =
\frac{10h_0^2}{27X^4}\frac{\pa^2 Q}{\pa X^2} + \left(X-1\right) \left( \frac{\pa^2  a}{\pa S^2} +\frac{4}{3X^3}\frac{\pa a}{\pa S}\right) \;,
\end{equation}
where
\begin{equation}
h_0^2 = \left[\frac{\dot{m}_0^2x_{\rm in}}{\sqrt{2}\eta m\alpha_0} \frac{\left(X-1\right)^2}{X} \right]^{1/5} \;.
\end{equation}
In geometrically thin accretion flows, the height satisfies $h_0\ll 1$. The terms proportional to $\pa^2Q/\pa S^2$ and $\pa^2a/\pa S^2$ in Eq.~\eqref{eq:pertaux2SS} can be neglected when the characteristic timescale of the fluctuations is comparable to the local viscous timescale, which is of the order of $X^3/h_0^2$ in our units. Neglecting these terms, and considering radii far from the inner radius of the accretion flow (i.e.,~$X\gg 1$), Eq.~\eqref{eq:pertaux2SS} can be approximated as 
\begin{equation}
\label{eq:pertfinalSS}
\frac{\pa Q}{\pa S} = \frac{1}{C_{\rm SS}X^{4/5}} \frac{\pa^2 Q}{\pa X^2} + \frac{4X}{3} \frac{\pa a}{\pa S} \;,
\end{equation}
where we defined
\begin{equation}
C_{\rm SS} = \frac{27}{10} \left(\frac{\sqrt{2}\eta m\alpha_0}{\dot{m}_0^2x_{\rm in}} \right)^{1/5} \;.
\end{equation}
For $m=10$, $\alpha_0=0.1$, $\dot{m}_0=0.1$, $x_{\rm in}=2$, we have $C_{\rm SS} = 9\times 10^4$. Eq.~\eqref{eq:pertfinalSS} is a diffusion-type equation with a source term proportional to $\pa a/\pa S$ \citep[see also Sect.~3 of][]{Lyubarsky1997}.

\begin{figure}
\centering
\includegraphics[width=0.49\textwidth]{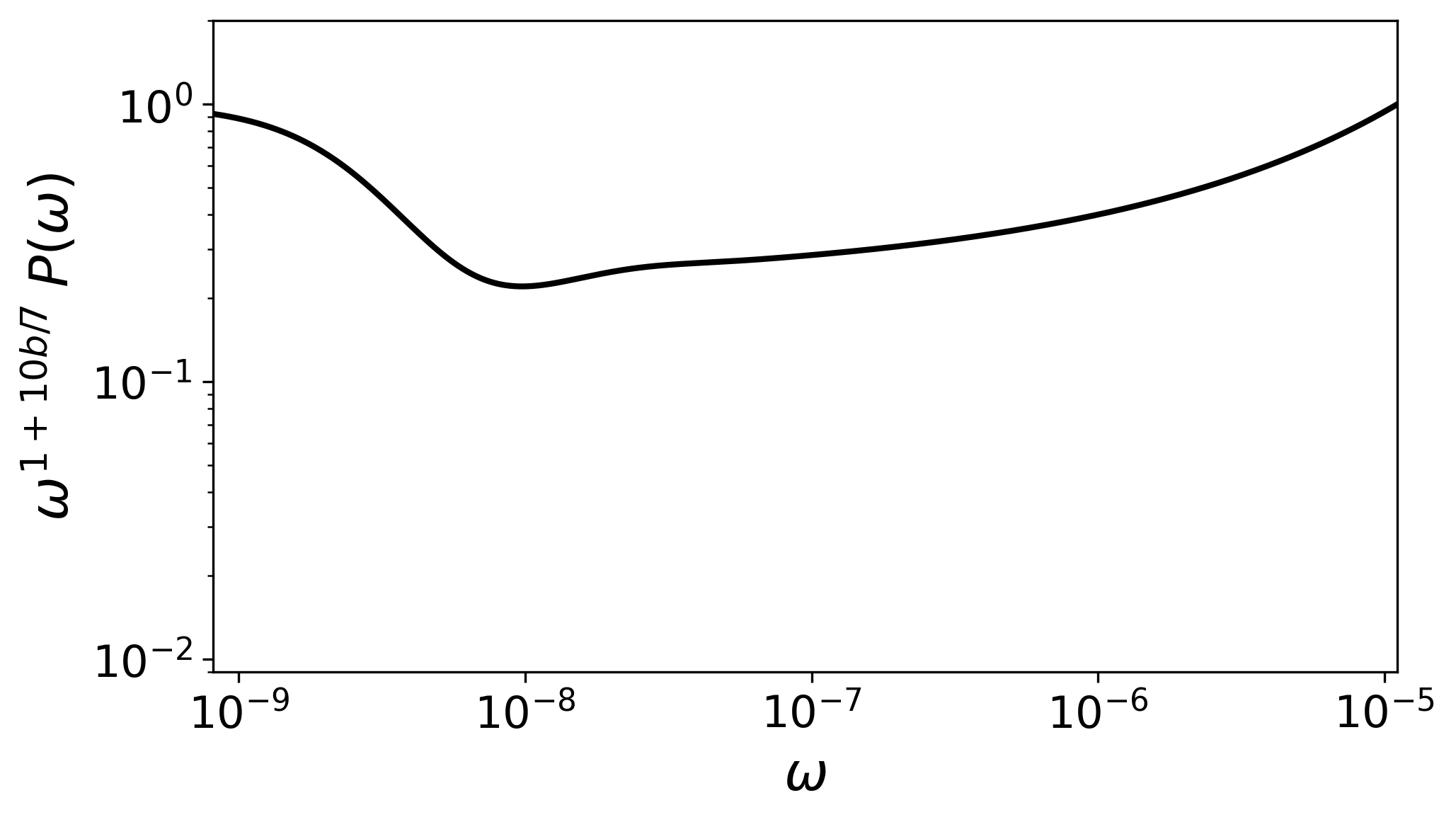}
\caption{\label{fig:powerSS}Power spectrum of the accretion rate onto the black hole for a geometrically thin cooling-dominated accretion flow. The power spectrum extends over the frequency range $C_{\rm SS}^{-1}X_{\rm out}^{-14/5}<\omega<C_{\rm SS}^{-1}$, where $C_{\rm SS} = 9\times 10^4$ and $X_{\rm out}=30$.}
\end{figure}

To calculate the power spectrum of the accretion rate, we follow the same procedure as in Sect.~\ref{sec:results}. Substituting Eqs.~\eqref{eq:fourier1}-\eqref{eq:fourier2} into Eq.~\eqref{eq:pertfinalSS}, we find
\begin{equation}
\label{eq:pertfinalSSfourier}
\left[{\rm i}\omega + \frac{1}{C_{\rm SS} X^{4/5}} \frac{\pa^2}{\pa X^2} \right]\tilde{Q} = \frac{4}{3} {\rm i}\omega X \tilde{a} \;.
\end{equation}
The homogeneous equation associated with Eq.~\eqref{eq:pertfinalSSfourier} is
\begin{equation}
\label{eq:homSS}
\frac{\pa^2\tilde{Q}}{\pa X^2} + {\rm i}\omega C_{\rm SS} X^{4/5} \tilde{Q} = 0 \;.
\end{equation}
The solution of Eq.~\eqref{eq:homSS} that satisfies the outer boundary condition at $X=X_{\rm out}$ is
\begin{align}
\nonumber
\tilde{Q}_{\rm out}\left(X,\omega\right) = & \; \sqrt{X}C_{J, {\rm out}}J_\nu\left[\frac{5}{7}\left(1+{\rm i}\right)\sqrt{\frac{C_{\rm SS}\omega}{2}}X^{7/5}\right] \\
& + \sqrt{X}C_{Y, {\rm out}}Y_\nu\left[\frac{5}{7}\left(1+{\rm i}\right)\sqrt{\frac{C_{\rm SS}\omega}{2}}X^{7/5}\right] \;,
\end{align}
where $J_\nu$ and $Y_\nu$ are the Bessel functions with $\nu=5/14$, and we defined
\begin{align}
C_{J, {\rm out}} & = Y_{\nu-1}\left[\frac{5}{7}\left(1+{\rm i}\right)\sqrt{\frac{C_{\rm SS}\omega}{2}}X_{\rm out}^{7/5}\right] \\
C_{Y, {\rm out}} & = - J_{\nu-1}\left[\frac{5}{7}\left(1+{\rm i}\right)\sqrt{\frac{C_{\rm SS}\omega}{2}}X_{\rm out}^{7/5}\right] \;.
\end{align}
The power spectrum of the accretion rate is given by
\begin{align}
\nonumber
P\left(\omega\right) =\int_1^{X_{\rm out}} & {\rm d}Y \int_1^{X_{\rm out}} {\rm d}Y' \frac{16}{9} C_{\rm SS}^2\omega^2 Y^{9/5}Y'^{9/5} \\
\label{eq:powerauxSS}
& \times \frac{\tilde{Q}_{\rm out}\left(Y,\omega\right) \tilde{Q}_{\rm out}^*\left(Y',\omega\right)}{\left|\tilde{Q}_{\rm out}\left(1,\omega\right)\right|^2} C_{\tilde{a}} \left(Y, Y', \omega\right) \;.
\end{align}
Since the local viscous timescale is of the order of $Y^3/h_0^2(Y)\sim C_{\rm SS}Y^{14/5}$, we assume that
\begin{equation}
\label{eq:corrSS}
C_{\tilde{a}} \left(Y, Y', \omega\right) \propto Y^{4b} \delta\left(\log Y-\log Y'\right) \delta\left(\omega-C_{\rm SS}^{-1}Y^{-14/5}\right) \;.
\end{equation}
Substituting Eq.~\eqref{eq:corrSS} into Eq.~\eqref{eq:powerauxSS}, we find
\begin{equation}
\label{eq:powerSS}
P\left(\omega\right) \propto \omega^{-1-10b/7} \left|\frac{\tilde{Q}_{\rm out}\left(C_{\rm SS}^{-5/14}\omega^{-5/14},\omega\right)}{\tilde{Q}_{\rm out}\left(1,\omega\right)}\right|^2 \;.
\end{equation}
Eq.~\eqref{eq:powerSS} gives the power spectrum of the accretion rate in the frequency range $C_{\rm SS}^{-1}X_{\rm out}^{-14/5}<\omega<C_{\rm SS}^{-1}$. The factor $\omega^{-1-10b/7}$ shows that the power spectrum depends on the frequency as a power-law \citep[see also Sect.~4 of][]{Lyubarsky1997}. The remaining factor does not generate QPOs, as shown in Fig.~\ref{fig:powerSS}.

\end{document}